\documentclass[reprint,amsmath,amssymb,aps,prl,superscriptaddress]{revtex4-2}

\usepackage{graphicx}

\usepackage{mathrsfs}
\usepackage{accents}
\usepackage{mystyle}
\usepackage{orcidlink}
\newcommand{\mytitle}{Hyperboloidal Initial Data with Prescribed Asymptotic Charges}
\usepackage{hyperref}
\hypersetup
{
  colorlinks=true,
  linktocpage=true,
  linkcolor={red},
  anchorcolor={black},
  citecolor={green},
  menucolor={red},
  unicode=true,
  pdftitle={\mytitle},
  pdfauthor={K. Z. Csukás, J. Östör, I. Rácz}
}

\definecolor{insC}{rgb}{0.0,0.45,0.45}   
\definecolor{ersC}{rgb}{0.55,0.0,0.75}   

\usepackage{amsthm}

\def\tf{\textit{\texttt{f}}}

\def\thf{\textit{\texttt{h}}}
\def\ta{\textit{\texttt{a}}}
\def\tb{\textit{\texttt{b}}}

\newcounter{mnotecount}

\newcommand{\mnotex}[1]
{
  \protect{\stepcounter{mnotecount}}
    $^{\mbox{\footnotesize $\bullet$\themnotecount}}$
    \marginpar{\color{red}\raggedright\tiny\em
  $\!\!\!\!\!\!\,\bullet$\themnotecount: #1}
}

\usepackage[normalem]{ulem}
\usepackage{color}
\definecolor{blue}{rgb}{0,0,1}
\definecolor{red}{rgb}{1,0,0}

\begin{document}

\title{\mytitle}

\author{Károly Csukás\,\orcidlink{0000-0002-2408-1103}}
\email{csukas.karoly@wigner.hu}
\affiliation{HUN-REN Wigner RCP, H-1121 Budapest, Konkoly Thege Mikl\'{o}s \'{u}t  29-33, Hungary}
\author{János Östör\,\orcidlink{0009-0000-5034-4088}}
\email{jostor@princeton.edu}
\affiliation{HUN-REN Wigner RCP, H-1121 Budapest, Konkoly Thege Mikl\'{o}s \'{u}t  29-33, Hungary}
\affiliation{Institute for Mathematics, Astrophysics and Particle Physics, Radboud University, 6525 AJ Nijmegen, The Netherlands}
\affiliation{Department of Physics, Princeton University, Princeton, New Jersey 08544, USA}
\author{István Rácz\,\orcidlink{0000-0002-1991-0472}}
\email{racz.istvan@wigner.hu}
\affiliation{HUN-REN Wigner RCP, H-1121 Budapest, Konkoly Thege Mikl\'{o}s \'{u}t  29-33, Hungary}

\date{\today}

\begin{abstract}
A novel method is introduced for constructing hyperboloidal initial data with direct analytic control over asymptotic charges prior to solving the Einstein constraint equations. Using the parabolic-hyperbolic formulation of the constraints together with a Fuchsian-analysis-inspired spectral solver, the asymptotic degrees of freedom, which determine infinite towers of asymptotic charges, including the Bondi energy and angular momentum, are used to prescribe the asymptotic behavior of solutions. As a proof of concept, we explicitly construct hyperboloidal initial data for a spinning black hole with Bondi energy $E_B = 2$ and angular momentum $J_B = 1/2$, starting from seed data induced on a hyperboloidal slice of the non-rotating Schwarzschild spacetime with 
mass $M=1$.
\end{abstract}

\maketitle

\section{Introduction}

The concept of self-gravitating isolated systems is an idealization. They may consist of stars and black holes, sufficiently remote from other sources. By introducing the notion of asymptotically simple spacetimes, Penrose provided a powerful mathematical framework to study such systems \cite{Penrose:1962ij, Frauendiener:2000mk, Kroon:2016ink}. In this setting, a fictitious boundary is attached to the physical spacetime, referred to as null infinity $\mathscr{I}$, which is represented by the ordinary boundary of a conformally related unphysical spacetime. Asymptotic simplicity requires that every inextendible null geodesic acquire past and future endpoints on $\mathscr{I}$. To accommodate black-hole spacetimes, which are among our principal targets throughout this paper, we will use the wider class of \emph{weakly} asymptotically simple spacetimes, which possess only a neighborhood of null infinity isometric to that of an asymptotically simple spacetime \cite{Kroon:2016ink,tod2021conformal}.

Penrose's construction assumes that the conformally rescaled metric is sufficiently smooth, which is essential for studying the falloff behavior of physical fields and, in particular, for deriving the peeling properties of the Weyl tensor \cite{Penrose:1962ij,Friedrich:1991nn,Friedrich:2015bax}. Building on this framework, the hyperboloidal initial value problem has emerged as an effective tool for constructing and analyzing weakly asymptotically simple spacetimes \cite{Friedrich1985,Friedrich:1991nn,Friedrich1996,Friedrich:2015bax,Frauendiener:2000mk,Kroon:2016ink}. Using his regular conformal field equations, Friedrich showed that smooth hyperboloidal data evolve into spacetimes admitting a smooth conformally regular future null infinity \cite{Friedrich:1991nn,Friedrich:2015bax}. 

All of the above theoretical developments, as well as many others on hyperbolic formulations of the Einstein equations, are related to, and to some extent rest on, the seminal results of Choquet-Bruhat and Geroch \cite{foures1952theoreme,bruhat2009general,choquet1969global}. These results guarantee, first, the existence and uniqueness of solutions to the evolutionary part of the Einstein equations, second, the existence of a maximal Cauchy development, and third, that the constraint equations hold throughout the domain of existence provided they are satisfied on the initial data surface. 

It follows that, prior to solving the Cauchy problem and determining the domain of dependence of the dynamical variables, suitable initial data must be prescribed. In the conventional Cauchy problem this is accomplished by solving the Einstein constraint equations on a spacelike hypersurface. In the hyperboloidal initial value problem, this initial data surface connects a spherical cut of null infinity $\mathscr{I}$ of the weakly asymptotically simple spacetime with a suitable interior region. For decades, the conformal (or elliptic) method introduced by Lichnerowicz and York \cite{Lichnerowicz_1944,York_1972,york1974covariant} was the only systematic approach to solving the constraints in Einstein's theory. This formulation offers straightforward control over the asymptotic decay of the constrained fields, yet, due to the conformal rescalings involved, it almost completely lacks control over distinguished asymptotic charges.

In the early 1990s, Andersson and Chruściel showed that the generic solutions of the elliptic form of the constraints in the hyperboloidal setting admit polyhomogeneous expansions near infinity \cite{Andersson:1992yk,Andersson:1993we,Andersson:1994ng,Andersson:1996xd}. More precisely, the asymptotic expansions of the constrained fields were found to involve not only integer powers of the conformal factor $\omega$ but also products of such terms and the logarithms of the conformal factor. They showed that the elimination of these logarithmic terms requires strong restrictions on the conformally rescaled extrinsic curvature at null infinity  \cite{Andersson:1992yk,Andersson:1993we,Andersson:1994ng,Andersson:1996xd}. Importantly, since logarithmic terms in the initial data propagate into the evolving spacetime metric \cite{Andersson:1996xd}, the desired smooth conformal decomposition of the physical metric may not be possible.

Two alternative {\it evolutionary} formulations of the constraint equations were introduced by one of us (IR), whereby the constraints are evolved in a spatial direction \cite{racz2015constraints}. Of these, the parabolic-hyperbolic formulation is particularly well suited to the hyperboloidal setting. Since this system is solved as an initial value problem, one might expect it to be very difficult, if not impossible, to control the asymptotic properties of the solutions. Nevertheless, it turns out that weakly asymptotically simple initial data can be obtained within the hyperboloidal setting simply by controlling the freely specifiable part of the initial data \cite{beyer2017asymptotics,beyer2020asymptotically,csukas2023possible}.

Beyer and Ritchie in \cite{Beyer:2021kmi} uncovered further favorable properties of the parabolic-hyperbolic formulation of the constraint equations. Using a Fuchsian-analysis-based argument \cite{kichenassamy2007fuchsian}, they proved that solutions to this formulation of the constraints are not only suited to the hyperboloidal initial value problem but also extend smoothly to the boundary at null infinity, provided that certain finite-order derivatives extend continuously there. Consequently, all derivatives of such solutions extend smoothly to $\mathscr{I}$ to arbitrary order, and no logarithmic terms appear in their asymptotic expansions.

The result of Beyer and Ritchie \cite{Beyer:2021kmi}, however, left open one of the most important questions: what physical requirements, if any, could guarantee the finite-order regularity of the constrained fields explicitly used in their Fuchsian-analysis-based argument? This became the primary motivation for our investigation in \cite{csukas-racz-2025-hyperboloidal}. We proved, while also slightly generalizing the analytic setting of \cite{Beyer:2021kmi}, that the requisite finite-order regularity holds precisely for those solutions of the parabolic-hyperbolic form of the constraint equations for which the Bondi energy and angular momentum are well-defined and finite \cite{csukas-racz-2025-hyperboloidal}.

The present paper grows out of the ideas developed in \cite{csukas-racz-2025-hyperboloidal}. Having gained control over the asymptotic degrees of freedom, it is natural to ask whether one can also a priori prescribe the asymptotic charges of hyperboloidal initial data constructed via the parabolic-hyperbolic formulation of the constraints. As the main result of this paper, we introduce a novel method for constructing hyperboloidal initial data with prescribed Bondi energy and angular momentum. In particular, we construct initial data for a non-stationary spinning black hole with Bondi energy $E_B = 2$ and angular momentum $J_B[\partial_\varphi] = 1/2$, from seed data induced on a hyperboloidal slice of the non-rotating Schwarzschild spacetime with ADM mass $M=1$.

  \section{The analytic setting}

  Vacuum initial data in general relativity consists of a triplet $(\Sigma,h_{ij},K_{ij})$, where $\Sigma$ is a three-dimensional manifold and $h_{ij}$ and $K_{ij}$ are symmetric tensor fields. These fields are not freely specifiable as they are subject to the constraint equations \cite{foures1952theoreme,Lichnerowicz_1944,York_1972,wald1984rg,choquet1969global}
  \begin{align}
  	\Rthree-K_{ij}K^{ij}+K^2&=0\,,\label{eq:Ham}\\
  	D_jK^j{}_i-D_iK&=0\,,\label{eq:Mom}
  \end{align}
  where $D_i$ is the covariant derivative compatible with $h_{ij}$, $\Rthree$ is the associated Ricci scalar, and $K=h^{ij}K_{ij}$ is the trace of $K_{ij}$. When a spacetime embedding of $\Sigma$ exists, $h_{ij}$ and $K_{ij}$ are referred to as the induced metric on $\Sigma$ and its extrinsic curvature $K_{ij}=\tfrac12\,\mathscr{L}_n h_{ij}$, respectively.
  
  When $\Sigma$ can be foliated by a one-parameter family of topological two-spheres $\mathcal{S}_\rho$, a $2+1$ split allows us to represent  $h_{ij}$ by the lapse $\Nhat$, shift $\Nhat^i$, and the induced metric $\gammahat_{ij}$, and $K_{ij}$ as scalar, vector, and tensor projections $(\kkappa,\kk_i,\KK_{ij})$. By further splitting the tensorial projection $\KK_{ij}$ into its trace $\KK$ and trace-free part $\Kc_{ij}$ as in \cite{racz2015constraints,racz2018toward}, and contracting all vector and tensor parts with the complex dyad $q_i=(d\vartheta)_i+\mathbf{i}\,\sin\vartheta\,(d\varphi)_i$, we obtain spin-weighted variables $(\Nhat,\NN,\aa,\bb;\kkappa,\kk,\KK,\Kcqq)$ adapted to the foliation $\mathcal{S}_\rho$, where $\NN$, $\bb$, $\kk$, $\Kcqq$ are complex (for details, see the End Matter \cite{SMtoMain}). The principal advantage of these variables is that angular differentiation reduces to an algebraic operation via the Newman--Penrose $\eth$ and $\ethb$ operators. In these variables, \eqref{eq:Ham} takes the form of a parabolic equation for $\Nhat$, while \eqref{eq:Mom} gives rise to a symmetric hyperbolic system for $\kk$ and $\KK$. This parabolic-hyperbolic system admits a locally well-posed initial value problem, which can be solved once the free data $(\NN,\aa,\bb;\kkappa,\Kcqq)$ on $\Sigma$ and initial data $(\Nhat|_{\rho_0},\kk|_{\rho_0},\KK|_{\rho_0})$ on one of the level sets $\mathcal{S}_{\rho_0}$ are specified.
  
  An initial data set $(\Nhat,\NN,\aa,\bb;\kkappa,\kk,\KK,\Kcqq)$ is then asymptotically hyperboloidal if $\lim_{\rho\rightarrow\infty}(\kkappa+\KK)>0$. When solving the constraints, we assume that the free data extends smoothly to $\partial\Sigma$, which is the intersection of $\Sigma$ and future null infinity. Three remarks justify singling out this class. First, on the initial slice the assumption is a choice rather than a restriction imposed by the field equations: the free data are, by definition, at our disposal, whereas the regularity of the constrained fields $(\Nhat,\kk,\KK)$, the subject of the results below, cannot simply be prescribed. Second, the smooth class is distinguished by the structure of the theory. As recalled above, generic hyperboloidal data are merely polyhomogeneous, and the logarithmic terms in their asymptotic expansions propagate into the evolving spacetime metric \cite{Andersson:1992yk,Andersson:1993we,Andersson:1994ng,Andersson:1996xd}, obstructing the smooth conformal compactification underlying Penrose's framework and the peeling properties of the Weyl tensor. Third, concerning the time development, Friedrich's semiglobal existence results \cite{Friedrich:1991nn,Friedrich:2015bax} guarantee that smooth hyperboloidal data evolve into spacetimes admitting a smooth conformally regular $\mathscr{I}^+$, at least in a neighborhood of the initial slice. Whether such smoothness persists globally for large data is a deep open problem lying beyond the scope of an initial-data construction. Restricting to smooth free data ensures that any loss of asymptotic regularity during evolution is not an artifact already planted in the initial data. Our main results in \cite{csukas-racz-2025-hyperboloidal,Racz:2024giv} rest on the use of the Bondi energy, which is the $\rho\to\infty$ limit of the Hawking energy, and angular momentum
  \begin{align}
  	& E_B =\lim_{\rho\to\infty}E_H=\lim_{\rho\to\infty}\sqrt{\frac{\mathcal{A}_\rho}{16\pi}}\left(1+\frac{1}{16\pi}\int_{\mathcal{S}_\rho}\theta^{(+)}\theta^{(-)}\,\epshat\right)\,,
  	\label{eq:Bondi-energy-0} \\
  	& J_B[\partial_\varphi] = - (8\, \pi)^{-1}\lim_{\rho\to\infty}\int_{{\mathcal{S}}_\rho} (\partial_\varphi)^i{ \rm \bf k}_i \, \widehat{{\boldsymbol{\epsilon}}} \,, \label{eq:Bondi-angmom-0}
  \end{align}
  where $\theta^{(\pm)}=\KK\pm\Kstar\Nhat^{-1}$ are the null expansions with $\Kstar=\tfrac12\gammahat^{ij}\mathscr{L}_{\rho-\Nhat}\gammahat_{ij}$, $\epshat$ is the volume element of $\gammahat_{ij}$, conformal to the volume element $\boldsymbol{\epsilon}_q=\sin\vartheta\, \mathrm{d}\vartheta\wedge\mathrm{d}\varphi$ of the center-of-mass unit sphere metric $q_{ij}=q_{(i}q_{j)}$, $\mathcal{A}_\rho=\int_{\mathcal{S}_\rho}\epshat$, and $\partial_\varphi$ is an axial vector field on the boundary  $\partial\Sigma\sim \mathbb{S}^2$ maximizing the right-hand side of \eqref{eq:Bondi-angmom-0} \cite{Racz:2024giv}. 
  In particular, in \cite{csukas-racz-2025-hyperboloidal,Racz:2024giv} we proved that the finiteness of the Bondi energy and angular momentum implies that the free data extends smoothly to $\partial\Sigma$ as
  \begin{align}
  	\NN(\omega,p)&=\NN_1(p)\,\omega+\NN_2(p)\,\omega^2+\mathcal{O}(\omega^3),\label{eq:bcNN}\\
  	\aa(\omega,p)&=\omega^{-2}+\aa_{(-1)}(p)\,\omega^{-1}+\aa_0(p)+\mathcal{O}(\omega),\\
  	\bb(\omega,p)&=\bb_0(p)+\bb_1(p)\,\omega+\mathcal{O}(\omega^2),\\
  	\kkappa(\omega,p)&=\kkappa_0(p)+\kkappa_2(p)\,\omega^2+\mathcal{O}(\omega^3),\\
  	\Kcqq(\omega,p)&=\Kcqq{}_0(p)+\Kcqq{}_1(p)\,\omega+\mathcal{O}(\omega^2),\label{eq:bcKcqq}
  \end{align}
   with $\aa_{(-1)}=\mathrm{const}$ and $\Kcqq{}_0=\frac{1}{2}\kkappa_0\,\eth\eth(\kkappa_0^{-2})$, where we use the parametrization $\omega=1/\rho$ so that $\omega=0$ represents the boundary $\partial\Sigma$ at infinity. Moreover, when $(\Nhat,\kk,\KK)$ satisfy the parabolic-hyperbolic form of the constraints, the finite-order expansions
  \begin{align}
  	\Nhat(\omega,p)&=\sum_{i=0}^4\Nhat_i(p)\,\omega^{i}+w_{\Nhat}(\omega,p)\,\omega^4\,,\label{eq:Nhseries}\\
  	\KK(\omega,p)&=\sum_{i=0}^{4}\KK_i(p)\,\omega^{i}+w_{\KK}(\omega,p)\,\omega^4\,,\label{eq:Kseries}\\
  	\kk(\omega,p)&=\sum_{i=0}^{3}\kk_i(p)\,\omega^{i}+w_{\kk}(\omega,p)\,\omega^3\,,\label{eq:kseries}
  \end{align}
  also hold. All coefficients depend on the angular coordinates $p\equiv(\vartheta,\varphi)$, while the remainder terms $w_{\Nhat},w_{\KK},w_{\kk}$ vanish on the boundary $\partial\Sigma$ at infinity.
 
  Remarkably, the parabolic-hyperbolic equations place no restriction on the three coefficients $(\KK_1, \kk_2, \Nhat_4)$, which we refer to as {\it the asymptotic degrees of freedom}. All other coefficients in the asymptotic expansions \eqref{eq:Nhseries}--\eqref{eq:kseries} are uniquely determined by the free data and $(\KK_1, \kk_2, \Nhat_4)$. Notably, although the asymptotic degrees of freedom $(\KK_1, \kk_2, \Nhat_4)$ are fields on the boundary $\partial\Sigma$ at infinity, they carry exactly the same information as initial data for the parabolic-hyperbolic form of the constraints specified on a regular $\omega=\omega_0$ level set.
  
  Since the asymptotic degrees of freedom $(\KK_1, \kk_2, \Nhat_4)$ are smooth functions on the boundary $\partial{\Sigma}$, and $\KK_1$ and $\Nhat_4$ are real, they admit expansions in spin-weight-zero spherical harmonics, while $\kk_2$ admits an expansion in spin-weight-one harmonics. The integrals \eqref{eq:Bondi-energy-0} and \eqref{eq:Bondi-angmom-0} [for explicit expressions in terms of the asymptotic degrees of freedom see \eqref{eq:B-energy-ADF} and \eqref{eq:Bondi-angmom-ADF}], however, depend only on the monopole parts of the integrands, yielding a substantial loss of information. This can be compensated for by introducing infinite towers of higher-mode Bondi energies $(E_B)_\ell{}^m$ and angular momenta $(J_B[\partial_\varphi])_\ell{}^m$, obtained by multiplying the integrands in \eqref{eq:Bondi-energy-0} and \eqref{eq:Bondi-angmom-0} by the complex conjugate $\overline{Y_\ell{}^m}$ of the spin-weight-zero spherical harmonics $Y_\ell{}^m$ and evaluating the resulting integrals. Notably, the $(\ell=1, m=0,\pm 1)$ modes of $(E_B)_\ell{}^m$ are related to the Bondi momentum \cite{szabados2004quasi}, but the physical interpretation of the full towers of these asymptotic charges remains to be explored. It would also be of considerable interest to determine how these quantities are related to the supermomentum and superspin charges that have attracted significant attention in recent studies of null infinity in weakly asymptotically simple spacetimes  \cite{barnich2010symmetries,flanagan2017conserved,elhashash2021definitions}.
 
  \section{Model and methodology}  
  
  It can be verified by direct computation that whenever the asymptotic expansions \eqref{eq:bcNN}--\eqref{eq:bcKcqq} and \eqref{eq:Nhseries}--\eqref{eq:kseries} hold, the constrained fields $(\Nhat,\kk,\KK)$ satisfy the parabolic-hyperbolic form of the constraints if and only if the vector-valued remainder $\underline{W}=\big(w_\KK(\omega,p),w_\kk(\omega,p),w_{\Nhat}(\omega,p)\big)$, with $\omega\in[0,\omega_0)$ and $p\in \mathbb{S}^2$, satisfies the Fuchsian-type system
  \begin{align}
  	\hskip-0.23cm \partial_\omega \underline{W} = {}&  \mathrm{diag}\big[-3,-1,0\big]\times(\omega^{-1} \underline{W}) \nonumber\\ &+ \underline{H}\big(\omega,p; \KK_1,\kk_2,\Nhat_4,\underline{W},\eth\underline{W},{\bar\eth}\underline{W},\eth{\bar\eth}\underline{W}\big)\,,\label{eq:Fuchsian1}
  \end{align}
  where $\underline{H}$ is a smooth, explicitly known vector-valued function that extends regularly to $\omega=0$. It is this equation that underlies the proof that the constrained fields $(\Nhat,\kk,\KK)$ extend smoothly to $\partial\Sigma$ whenever the Bondi energy and angular momentum are finite \cite{Beyer:2021kmi,csukas-racz-2025-hyperboloidal}.
  
  Recall that our principal aim is to construct solutions of the constraint equations with prescribed Bondi energy and angular momentum. Since the asymptotic charges \eqref{eq:Bondi-energy-0} and \eqref{eq:Bondi-angmom-0} depend explicitly on the free data and on the asymptotic degrees of freedom $(\KK_1, \kk_2, \Nhat_4)$, and since precisely these quantities also appear explicitly in the Fuchsian equation \eqref{eq:Fuchsian1} \cite{Beyer:2021kmi,csukas-racz-2025-hyperboloidal}, they offer a direct means of controlling the asymptotic charges.
  
  To obtain solutions of the desired type, the most natural approach would be to integrate \eqref{eq:Fuchsian1} towards $\partial\Sigma$ starting from initial data $(\Nhat|_{\omega_0}, \kk|_{\omega_0}, \KK|_{\omega_0})$ prescribed on the level set $\omega = \omega_0>0$. Solutions obtained in this way, however, tend to develop singularities at $\omega = 0$, because both the initial data $(\Nhat|_{\omega_0}, \kk|_{\omega_0}, \KK|_{\omega_0})$ and the asymptotic degrees of freedom $(\KK_1, \kk_2, \Nhat_4)$ at $\omega = 0$ enter as independent input, rendering the problem overdetermined. An alternative would be to integrate \eqref{eq:Fuchsian1} for the remainder terms $w_{\KK}, w_{\kk}, w_{\Nhat}$ inward from $\partial\Sigma$, exploiting the fact that these remainders vanish at $\omega = 0$. This approach, however, conflicts with the natural direction of integration for the parabolic equation involved, causing a loss of angular spectral convergence.
  
  As discussed above, when the Bondi energy and angular momentum are well-defined and the parabolic-hyperbolic form of the constraints is satisfied, the constrained fields $(\Nhat,\kk,\KK)$ are guaranteed to be regular to finite order. Moreover, the Fuchsian equation \eqref{eq:Fuchsian1} implies that derivatives of $(\Nhat,\kk,\KK)$ of arbitrary order extend continuously to $\partial\Sigma$ \cite{Beyer:2021kmi,csukas-racz-2025-hyperboloidal}.
    
 In light of these observations, our main objective is to find the unique regular solution to the Fuchsian system \eqref{eq:Fuchsian1}, which, in particular, must vanish on $\mathscr{I}$ and its $\omega$-derivatives up to arbitrary order are also known there. Pseudo-spectral methods are known to select such a solution effectively. In order to find it, we first expand the vector-valued variable $\underline{W}=\big(w_\KK,w_\kk,w_{\Nhat}\big)$ in terms of spin-weighted spherical harmonics. The $\omega$-dependence of the expansion coefficients is then determined using a pseudo-spectral code based on Chebyshev--Lobatto nodes. To obtain a sufficiently good approximation to the desired regular solution, we minimize the residuals of the expansion coefficients at the $N_\omega$ Chebyshev--Lobatto collocation points associated with our Fuchsian system \eqref{eq:Fuchsian1} and the vector-valued variable $\underline{W}$ (discussed in more detail in the End Matter). The cutoff $\ell_{\max}$ in the spin-weighted spherical harmonics and the number $N_\omega+1$ of collocation points were then chosen, by inspection, so that the combined $L^2$-norm of the residuals remained below the desired tolerance over the entire horizon-penetrating initial data surface.
  
  \section{Spinning up a black hole}
  
\begin{figure}[h]
  	\includegraphics[width=.45\textwidth]{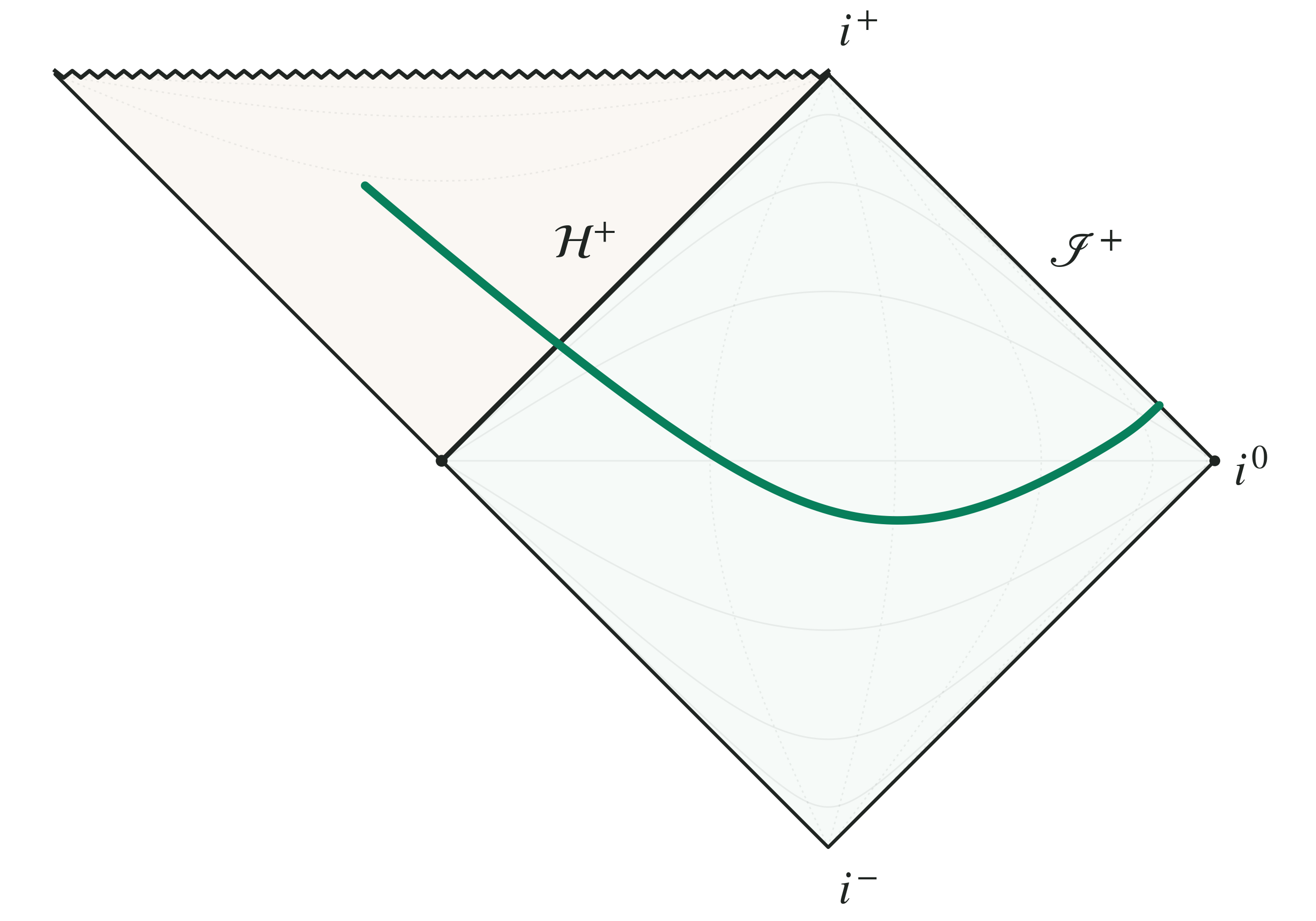}
  	\vskip-0.2cm \caption{The initial $T=\mathrm{const}$ hyperboloidal slice of the background Schwarzschild spacetime, determined by the height function (see the End Matter for details), shown on a Carter--Penrose diagram. This slice intersects both future null infinity $\mathscr{I}^{+}$ and the future event horizon $\mathcal{H}^+$ in regular cuts.}
  	\label{fig:Tconst}
  \end{figure}

  To demonstrate the viability of the proposed method we take the free data to be those induced on a $T=\mathrm{const}$ hyperboloidal slice
  of a non-rotating Schwarzschild spacetime of mass $M$, determined by the relation $t=T+\thf(\rho)$, where $t$ and $\rho$ denote the standard Schwarzschild time and area-radial coordinates, and $\thf(\rho)$ is a suitable height function. 
  As discussed in detail in the End Matter, the explicit choice \eqref{eq:b-func} we made for $\tb$ also determines the height function $\thf(\rho)$. It is worth emphasizing here that the $T=\mathrm{const}$ slice connects future null infinity to the future event horizon and remains well-defined as it extends deep into the black-hole region, as depicted in Fig.~\ref{fig:Tconst}.

   For such a solution, one has
  \begin{equation}
  	\aa_1=\aa_0=\NN_1=\NN_2=\eth\kkappa_0=0\,.
  \end{equation}
  Inserting these expressions, along with a choice of asymptotic degrees of freedom $(\KK_1, \kk_2, \Nhat_4)$, into \eqref{eq:Bondi-energy-0} and \eqref{eq:Bondi-angmom-0}, the Bondi energy and angular momentum reduce to
  \begin{align}
  	E_B = \frac{1}{32\pi}\int_{\partial{\Sigma}}\Big\{&\KK_1\big[\KK_1{}^2-6\big]{\kkappa_0}^{-1}+8\big[\KK_1\kkappa_2+\Nhat_4\,\kkappa_0{}^3\big] \nonumber\\
  	&-4\,\kkappa_0\,\kkappa_3 - \kkappa_0{}^{-1}\eth\ethb\KK_1\Big\}\,\boldsymbol{\epsilon}_q\,,\label{eq:Bondi-energy}\\
  	J_B[\partial_\varphi] = -\frac{1}{16\pi}&\int_{\partial{\Sigma}}\Big\{\big[(\partial_\varphi)^i\, \overline{q}_i\big]\,\kk_2 + \big[(\partial_\varphi)^i q_i\big]\,\overline{\kk}_2\Big\}\,\boldsymbol{\epsilon}_q\,.\label{eq:Bondi-angmom}
  \end{align}
  
  As noted above, the integrals in \eqref{eq:Bondi-energy} and \eqref{eq:Bondi-angmom} are sensitive only to the monopole parts of the integrands. Since the free data are spherically symmetric, all the $\kkappa_i$ terms are constant on $\partial\Sigma$. Moreover, as $\Nhat_4$ enters \eqref{eq:Bondi-energy} linearly, only its monopole part contributes to the Bondi energy. By contrast, since $\KK_1$ appears in the integrand of \eqref{eq:Bondi-energy} via a cubic polynomial, it is less straightforward to control the Bondi energy by varying $\KK_1$. 
  To demonstrate that $\KK_1$ and $\Nhat_4$ can be varied simultaneously to achieve the desired value of the Bondi energy ${E_B}= 2\,\bg{E_B}$, we proceed as follows. For simplicity, we set $M = 1$. The corresponding seed data, computed for the chosen time slice, are $\bg\KK_1\approx-93.3500$ and $\bg{\Nhat}_4\approx-6.4058$. These values are obtained by evaluating the background variables on the initial slice, which we also chose to provide favorable numerical properties for the second hyperboloidal slice, as discussed around \eqref{eq:intGfk} in the End Matter. Note that inserting the precise values of $\bg\KK_1$ and $\bg{\Nhat}_4$ into \eqref{eq:Bondi-energy} indeed reproduces $\bg{E_B} = M$. Because of the less straightforward dependence of the Bondi energy on $\KK_1$, we prescribe $\KK_1$ first, setting $\KK_1=\bg\KK_1+\alpha\cdot Y_0{}^0$ with $\alpha=-36.8911$. We then set $\Nhat_4 = \bg{\Nhat}_4 + \beta \cdot Y_0{}^0$, where $\beta$ is determined by the condition ${E_B}= 2\,\bg{E_B}$, which, by virtue of \eqref{eq:Bondi-energy} and the linearity of $E_B$ in $\Nhat_4$, is a linear equation for $\beta$. Solving it in closed form yields $\beta =291.041$, the truncation of the exact closed-form value. Note that no iteration is used here. The choice we made for the values of $\alpha$ and $\beta$ exactly double the Bondi energy of the background, giving $E_B = 2\,\bg{E_B} = 2$. The choices for $\alpha$ and $\beta$ were made to obtain a spherically symmetric, mass-two Schwarzschild companion, which is used below when adding nontrivial angular momentum.
  
  Analogously, replacing the asymptotic value $\bg{\kk}_2 = 0$ of the non-rotating Schwarzschild spacetime with $\kk_2 = \gamma \cdot {}_1Y_1{}^0$, where  
  $\gamma$ is purely imaginary, yields initial data whose Bondi angular momentum, obtained by direct evaluation of \eqref{eq:Bondi-angmom}, is $J_B[\partial_\varphi] = |\gamma|/\sqrt{24\pi}$. The closed-form choice $\gamma = -\sqrt{6\pi}\cdot{\bf i}$ therefore yields $J_B[\partial_\varphi] = 1/2$ exactly.
  Here $\partial_\varphi$ denotes the axial Killing vector field of the center-of-mass unit sphere metric on $\partial\Sigma$, whose axis is aligned with the ${}_1Y_1{}^0$ mode and which maximizes the functional \eqref{eq:Bondi-angmom}.  
  We emphasize that no numerical tuning or iteration is involved in prescribing the parameters $\alpha$, $\beta$, and $\gamma$. The values of $\alpha$ and $\beta$ quoted above depend on the choice we made for the mass-two companion slice that was selected solely to optimize the numerical accuracy.
  
  The proposed method represents the constrained fields $(\Nhat,\kk,\KK)$ on the secondary $T=\mathrm{const}$ hyperboloidal slice of the non-rotating Schwarzschild spacetime by combining a decomposition in spin-weighted spherical harmonics with a pseudo-spectral code. 
  The key question is whether one can find a moderate cutoff $\ell_{\max}$ in the spin-weighted harmonics, a modest number $N_\omega$ of Chebyshev--Lobatto collocation points, and a value $\omega_0>0$ such that the portion of the slice corresponding to $\omega\in[0,\omega_0]$ is horizon-penetrating and  $(\Nhat,\kk,\KK)$ satisfy the constraint equations to within the desired numerical tolerance. We find that choosing the parameters $\ell_{\max}=20$ and $N_\omega=192$ keeps the combined $L^2$-norm of the residuals below $10^{-24}$ over the entire interval $0\leq \omega\leq 0.3$. The obtained deformed black-hole initial data set thus satisfies the constraints up to a very high accuracy on a slice extending from $\mathscr{I}$ to well inside the black-hole region.
  
  Concerning the location of the marginally outer trapped surface (MOTS) on $\Sigma$, we emphasize that the constructed data are neither spherically symmetric nor stationary, so no exact relation ties the MOTS radius to the Bondi energy, especially since part of the energy resides in the gravitational field of the black-hole exterior. Nevertheless, the crude heuristic that doubling the Bondi energy of a Schwarzschild seed roughly doubles the horizon radius suggests searching for the MOTS near $\omega \approx 1/4$, safely in the interior region of the domain of validity $0 \leq \omega \leq \omega_0$. The MOTS located numerically is indeed close to, though measurably distinct from, this naive estimate. Its coordinate radius is $\rho = 3.9905$ on the equator $(\vartheta=\pi/2)$ and $\rho = 3.9953$ at the poles $(\vartheta=0, \pi)$, implying that $\omega_{\mathrm{mots}} \approx 0.2506 < \omega_0 = 0.3$. Accordingly, the MOTS is slightly shrunken (with respect to $\omega = 1/4$) and prolate, consistent with $E_{\mathrm{mots}}<E_B$. 
  
\begin{figure*}[t]
  	\includegraphics[width=.98\textwidth,height=7cm
  	]{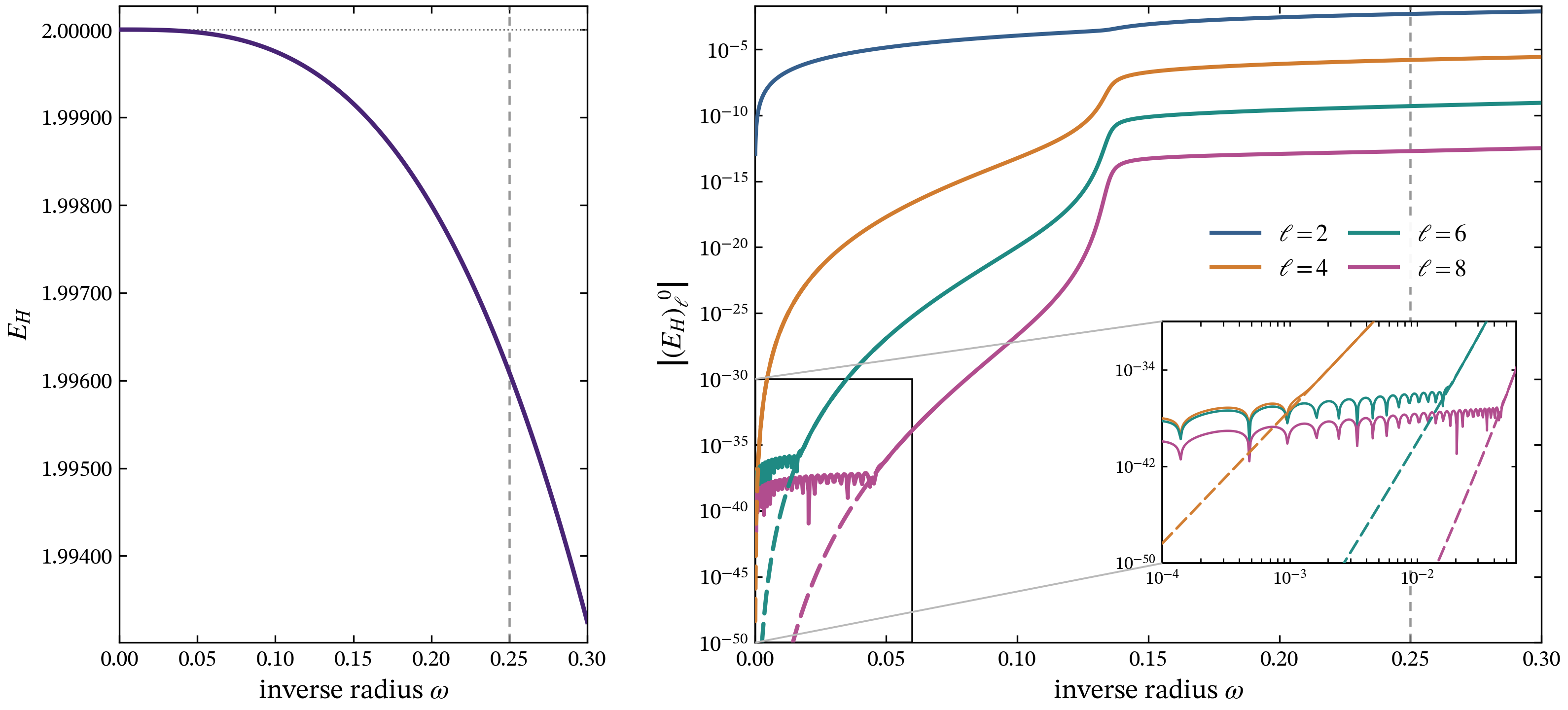}
  	\vskip-0.2cm \caption{In the left panel, the $\omega$-dependence of the Hawking energy $E_H=(E_H)_{0}{}^0$ is shown. In the right panel, the $\omega$-dependence of the first four even $\ell$-mode aspects $(E_H)_{\ell}{}^0$ of the Hawking energy is plotted. Solid lines show the direct numerical data obtained by spectral extraction, and dashed lines show their asymptotic behavior as derived from an order-$50$ Taylor series solution of the constraints. The inset depicts finite accuracy cancellation effects obtained from the direct pseudo-spectral extraction near $\mathscr{I}$.  The vertical line indicates $\omega=1/4$, which is slightly smaller than $\omega_{\mathrm{mots}}$. 
  	}
  	\label{fig:HE}
  \end{figure*}

  The deviation of $\rho_{\mathrm{mots}}$ from $\rho= 4$ reflects the energy content of the exterior region, which is quantified by the inward decrease of the Hawking energy shown in Fig.~\ref{fig:HE}. It takes the value $E_H = E_B = 2$ at the boundary $\partial\Sigma$ at infinity, represented by $\omega = 0$, and decreases slowly moving inward.
  
  Interestingly, not only the Bondi angular momentum $J_B[\partial_\varphi]=(J_B[\partial_\varphi])_{0}{}^{0}$, but also its ($\ell=2, m=0$) aspect $(J_B[\partial_\varphi])_{2}{}^{0}$ is excited. Nevertheless, neither of the corresponding quasi-local charges depend on the radial coordinate, as follows from (4.5) of \cite{Racz:2024giv} for axially symmetric initial data. Indeed, since each $\omega = \mathrm{const}$ level set on $\Sigma$ serves as a center-of-mass spherical reference surface, one can verify directly that $J[\partial_\varphi] = 1/2$ and $(J[\partial_\varphi])_2{}^{0}=-1/(2\sqrt{5})$ hold for these quasi-local charges independently of $\omega$. 
  
\begin{figure}[h]
  	\includegraphics[width=.48\textwidth,height=7.5cm]{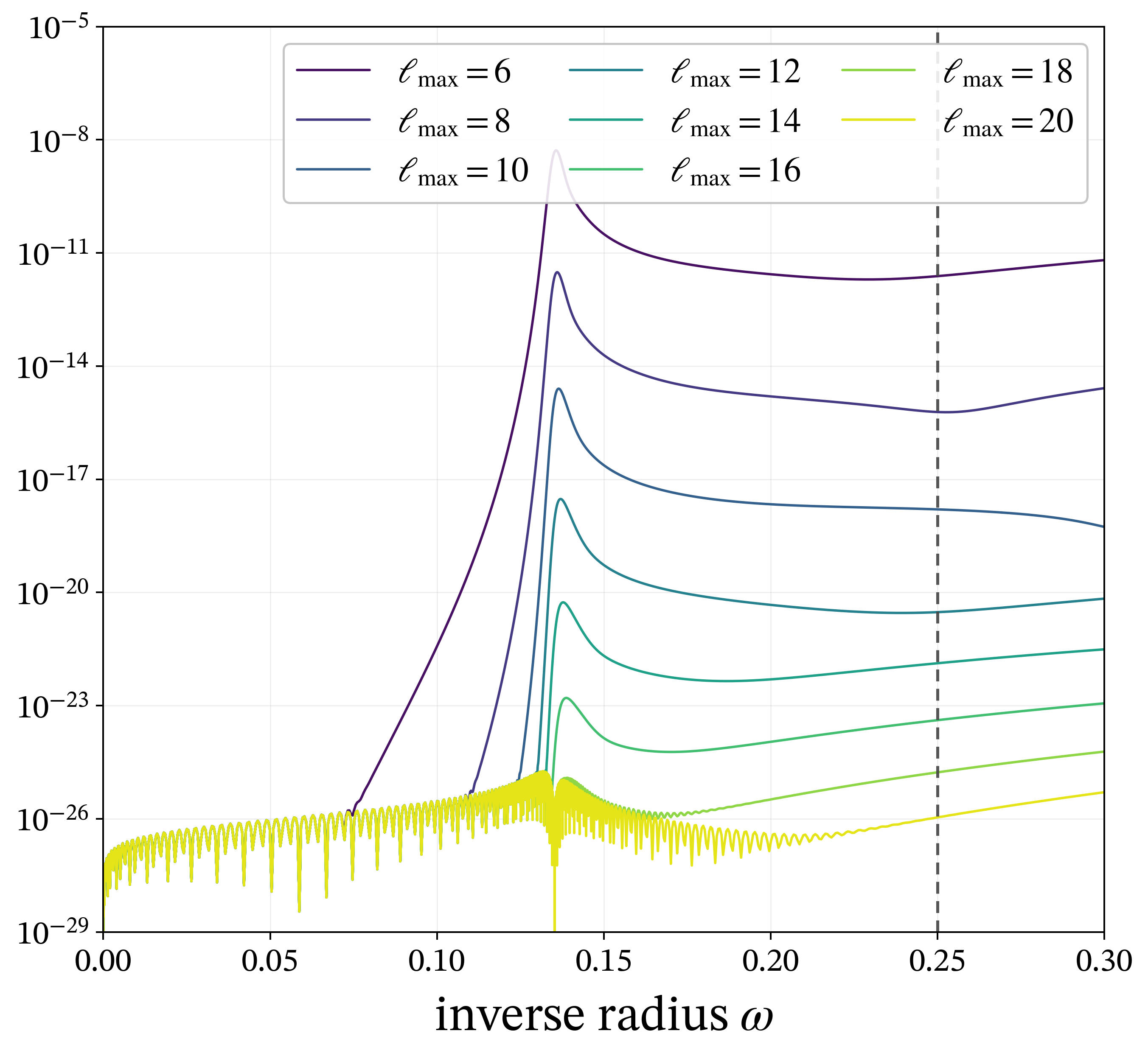}
  	\vskip-0.2cm 
  	\caption{The $\omega$-dependence of the combined $L^2$-norm of the residuals $\mathcal E_{\rm full}(\omega)$ (defined in the End Matter) is plotted for the fixed radial resolution $N_\omega=192$ of the Chebyshev--Lobatto collocation points, while varying $\ell_{\max}$ in the range $\ell_{\max}=6,8,\dots,20$. The angular convergence and the numerical accuracy achieved verify the effectiveness of our numerical method. As in Fig.~\ref{fig:HE}, the vertical dashed line indicates $\omega=1/4\lesssim\omega_{\mathrm{mots}}$.}
  	\label{fig:res}
  \end{figure}

\begin{figure}[h]
  	\includegraphics[width=.48\textwidth,height=7.5cm]{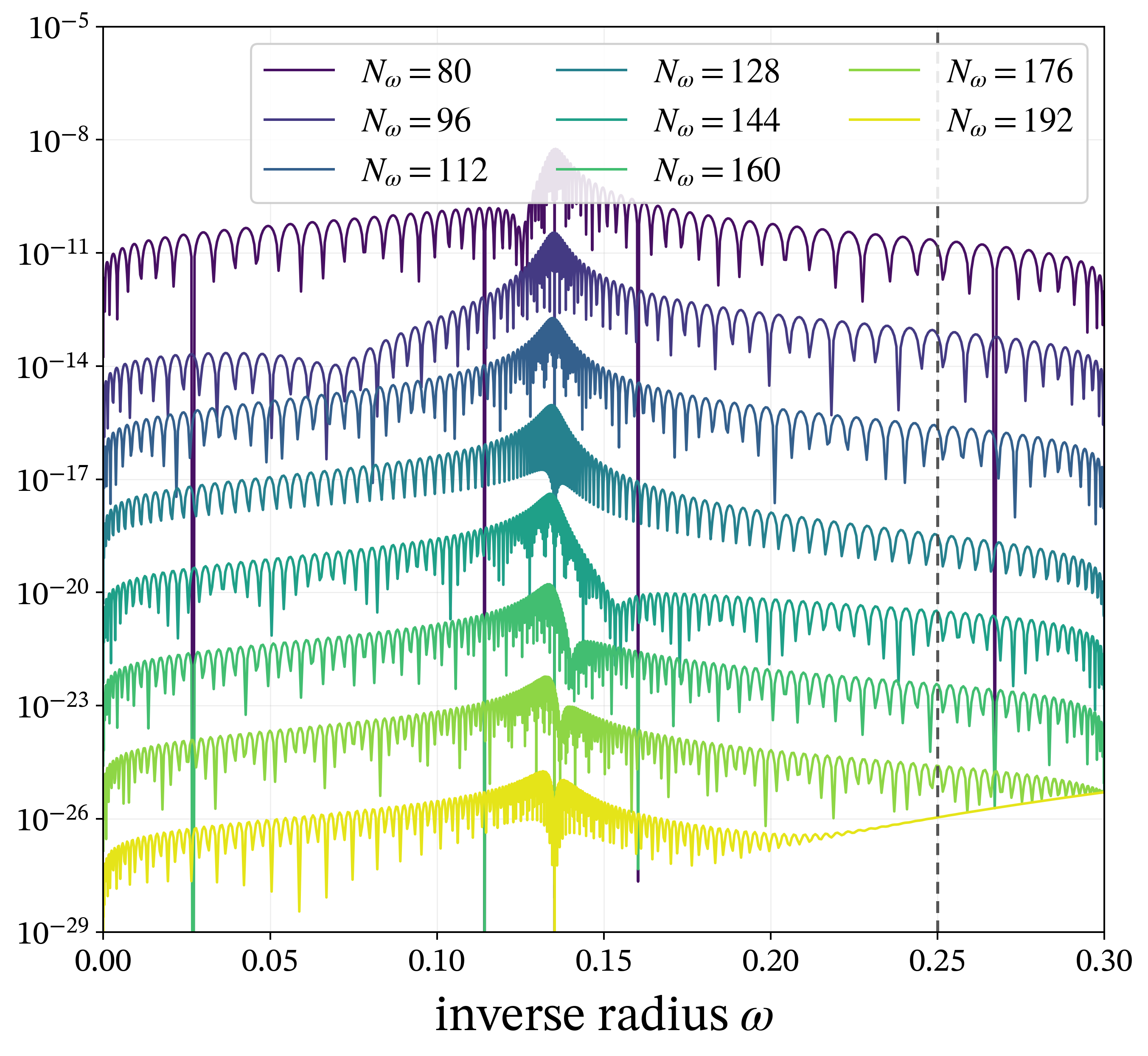}
  	\vskip-0.2cm 
  	\caption{The $\omega$-dependence of the combined $L^2$-norm of the residuals $\mathcal E_{\rm full}(\omega)$ (defined in the End Matter) is plotted for the fixed value $\ell_{\max}=20$, while varying the number of Chebyshev--Lobatto collocation points in the range $N_\omega=80,96,\dots,192$. The systematic downward shift of the curves confirms the expected radial convergence. As in Fig.~\ref{fig:HE}, the vertical dashed line indicates $\omega=1/4\lesssim\omega_{\mathrm{mots}}$.}
  	\label{fig:res2}
  \end{figure}

  To verify the viability of the proposed method, we performed several numerical checks. Figs.~\ref{fig:res} and~\ref{fig:res2} show the $\omega$-dependence of the combined $L^2$-norm of the constraint residuals associated with the constrained fields $(\Nhat,\kk,\KK)$.
  
  This combined $L^2$-norm remains below $10^{-24}$ for all $0 \leq \omega \leq \omega_0 = 0.3$ with the resolution $\ell_{\max}=20$ and $N_\omega=192$.  Figs.~\ref{fig:res} and~\ref{fig:res2} verify both the radial and the angular spectral convergence of our numerical method.
  
  \section{Conclusion}
  
  We have introduced a novel method for constructing vacuum hyperboloidal initial data with direct analytic control over asymptotic physical charges. By combining the parabolic-hyperbolic formulation of the Einstein constraints with a Fuchsian-analysis-inspired pseudo-spectral method, infinite towers of asymptotic observables, including the Bondi energy and angular momentum, can be prescribed prior to solving the constraints. As a proof of concept, we explicitly constructed highly accurate black hole initial data, representing solutions of the constraints on a hyperboloidal initial data surface extending from a spherical cut of null infinity to an inner boundary well inside the horizon.
  
  It is worth noting that hyperboloidal initial data for single and binary systems containing black holes and neutron stars have previously been constructed using the elliptic formulation of the constraints \cite{Buchman:2009ew,Bardeen:2011pd,Schinkel:2013zm,Vano-Vinuales:2025oai}. A significant limitation of these constructions, however, is that, apart from the Bondi energy-momentum in \cite{Bardeen:2011pd}, controlling the asymptotic charges requires an iterative search over the free parameters until a solution with the desired physical properties is found. This is precisely where the parabolic-hyperbolic approach offers a key advantage. It eliminates the trial-and-error iteration inherent in the elliptic method by providing direct control over infinite towers of asymptotic charges prior to solving the constraints.
  
\begin{acknowledgments}
  This work was supported in part by the Hungarian Scientific Research Fund NKFIH Grant No. K-142423. The authors are grateful to the Erwin Schrödinger Institute for its hospitality during the workshop ``Hyperboloidal Foliations and their Applications'', where preliminary results of this work were presented.
  
\end{acknowledgments}

\bibliography{asymhyper2.bib}



\appendix

\begin{widetext}
	\section*{End Matter}
\end{widetext}
\renewcommand{\thesection}{EM.\arabic{section}}
\renewcommand{\thesubsection}{EM.\arabic{subsection}}
\newtheorem{lemma}{Lemma}[subsection]
\renewcommand{\thelemma}{EM.\arabic{lemma}}
\renewcommand{\theequation}{EM.\arabic{equation}}
\setcounter{equation}{0}

	\section{The height function determining the $T=\mathrm{const}$ slice}

	As mentioned in the main text, the horizon-penetrating hyperboloidal $T=\mathrm{const}$ slices of the Schwarzschild spacetime of mass $M$ are constructed by means of a suitable height function, $t = T + \thf(\rho)$ \cite{zenginouglu2007conformal}. The height function is selected as follows. In the desired local coordinates
	$(T,\rho,\vartheta,\varphi)$, the background Schwarzschild metric takes the form
	\begin{equation}\label{eq:heightmetric}
		ds^2 = -\tf\, dT^2 - 2\,\tb\, dT\, d\rho + \ta\, d\rho^2
		+ \rho^2 \big( d\vartheta^2 + \sin^2\vartheta\, d\varphi^2 \big)\,,
	\end{equation}
	where $\tf = 1 - \tfrac{2M}{\rho}$, while the boost function $\tb$ and the radial metric function $\ta$ are related to the height function via $\tb = \tf\cdot\thf'$ and $\ta = \tf^{-1}\big(1-\tb^2\big)$. Note that on a $T=\mathrm{const}$ slice the lapse of the $\rho$-foliation is $\Nhat = \sqrt{\ta\,}$. These relations guide us in the choice of the initial $\thf$. The boundary behavior $\tb \to 1$ as $\rho \to \infty$ and $\tb \to -1$ as $\rho \to 2M$ guarantees that the $T=\mathrm{const}$ slices intersect both future null infinity $\mathscr{I}^{+}$ and the future event horizon of the background Schwarzschild spacetime in regular cuts.
	This requirement led us to choose
	\begin{equation}\label{eq:b-func}
		\tb(x)=\frac{e^{g(x)}\,(1-x)-e^{-g(x)}\,x^2+\lambda\,x}{\sqrt{[e^{g(x)}\,(1-x)-e^{-g(x)}\,x^2+\lambda\,x]^2+4\,x^2\,(1-x)}}
	\end{equation}
	where $x = 2M\omega = 2M\rho^{-1}$, $\lambda=0.1629$, and
	\begin{widetext}
		\begin{equation}\label{eq:g-func}
			g(x) = (1 - x)\, b_0 + x \,  b_1 + x (1 - x) \, b_2   +  x (1 - x) (2 x - 1)\, b_3   + x^2 (1 - x)^2 \, b_4  + \tfrac12\,A \Big[1 - \tanh\big(\tfrac{x - x_d}{w}\big)\Big]\,,
		\end{equation}
	\end{widetext}
	where the real parameters $b_0,b_1,b_2,b_3,b_4,x_d,A,w$ were fixed as $b_0=0.9439$, $b_1=0.0378$, $b_2=-2.0573$, $b_3=1.8144$, $b_4=-0.1497$, $x_d=0.29$, $A=1.8395$, $w=0.06$ in order to obtain, by integrating the relation $d\thf/d\rho=\tb/\tf$, the initial $T=\mathrm{const}$ slice depicted in Fig.~\ref{fig:Tconst}. Note that the resulting boost function $\tb$ also has the expected boundary behavior, i.e., $\tb = 1$ at $\omega = 0$ and $\tb = -1$ at $\omega = 1/(2M)$.
	
	The choice we made for the initial $T=\mathrm{const}$ slice of the background Schwarzschild spacetime, constructed as described above, was motivated by the following observations. First, note that this choice influences the spherical mass-two companion hyperboloidal slice,  
	introduced in the main text, which corresponds to the excitations $\KK_1=\bg\KK_1+\alpha\,Y_0{}^0$ and $\Nhat_4=\bg\Nhat_4+\beta\,Y_0{}^0$, and to Bondi energy $E_B=2$. Second, apart from intersecting both future null infinity $\mathscr{I}^{+}$ and the future event horizon in regular cuts, both the initial and the second slices also implicitly affect the achievable numerical accuracy. Indeed, the linear-perturbation argument given below suggests that minimizing the exponent $\mathfrak{G}(\omega)$ in \eqref{eq:intGfk} results in the second hyperboloidal slice suitable for achieving the desired numerical accuracy. This guiding principle was used in setting up our numerical method. 

\section{The spin-weighted variables}\label{app:1}

The spin-weighted variables $(\Nhat,\NN,\aa,\bb;\kkappa,\kk,\KK,\Kcqq)$, introduced briefly in the main text, are defined more precisely as follows.

Assume that the initial data surface $\Sigma$ can be foliated by a one-parameter family of topological two-spheres $\mathcal{S}_\rho$, which are level surfaces of some smooth real function $\rho:\Sigma\rightarrow\mathbb{R}$. The underlying $2+1$ decomposition allows us to define the unit normal to the $\rho=\mathrm{const}$ level surfaces as $\nhat_i=(d\rho)_i/\sqrt{h^{kl}(d\rho)_k(d\rho)_l}$. Using the three-metric $h_{ij}$ and its inverse, we define the projector  as $\gammahat_{i}{}^j=h_{i}{}^j-\nhat_i\nhat^j$, satisfying $\gammahat_i{}^{j}\nhat_{j}=0$ and $\gammahat_i{}^l\gammahat_l{}^j=\gammahat_i{}^j$. We also choose an evolution vector field $\rho^i$ on $\Sigma$ whose integral curves intersect each level surface exactly once, normalized so that $\rho^i\partial_i\rho=1$ \cite{racz2015constraints,racz2018toward}.

Analogously, we take the scalar, vector, and tensor projections of $K_{ij}$, defined respectively as $\kkappa=K_{ij}\nhat^i\nhat^j$, $\kk_i=\gammahat_i{}^j\nhat^lK_{jl}$, and $\KK_{ij}=\gammahat_i{}^k\gammahat_j{}^lK_{kl}$. It is also useful to decompose $\KK_{ij}$ into its trace $\KK=\gammahat^{ij}K_{ij}$ and trace-free part $\Kc_{ij}=\KK_{ij}-\frac{1}{2}\gammahat_{ij}\KK$. This yields an algebraically equivalent representation of the initial data $(h_{ij},K_{ij})$ via $(\Nhat,\Nhat^i,\gammahat_{ij};\kkappa,\kk_i,\KK,\Kc_{ij})$. All of the latter fields can be viewed as intrinsic to the $\rho=\mathrm{const}$ level sets.

Choosing standard spherical coordinates $(\vartheta,\varphi)$ on one of the $\rho=\mathrm{const}$ level surfaces, we define the complex null dyad $q_i=(d\vartheta)_i+\mathbf{i}\,\sin\vartheta\,(d\varphi)_i$, with contravariant form $q^i=(\partial_\vartheta)^i+\mathbf{i}\,(\sin\vartheta)^{-1}\,(\partial_\varphi)^i$, satisfying $q_iq^i=0$, $q_i\qbar^i=2$, and $q_{ij}=\frac{1}{2}(q_i\qbar_j+\qbar_iq_j)=\mathrm{d}\vartheta^2+\sin^2\hskip-2pt\vartheta\,\mathrm{d}\varphi^2$. The spherical coordinates $(\vartheta,\varphi)$ and the complex null dyad are then extended to all $\rho=\mathrm{const}$ level surfaces by Lie dragging along the evolution vector field $\rho^i$ \cite{racz2015constraints,racz2018toward}. Using these auxiliary fields, we evaluate the contractions of the dyad fields with the vector and tensor projections introduced above. The scalar variables $\Nhat$, $\kkappa$, $\KK$ are automatically of spin-weight zero and real, while the dyad components are $\NN=\Nhat_iq^i$, $\aa=\tfrac{1}{2}\gammahat_{ij}q^i\qbar^j$, $\bb=\frac{1}{2}\gammahat_{ij}q^iq^j$, $\kk=\kk_iq^i$, and $\Kcqq=\Kc_{ij}q^iq^j$.

\section{Spin-weighted harmonic representation}

The constrained variables ${\bf K}$ and ${\widehat N}$ are of spin weight $s=0$, whereas ${\bf k}$ is of spin weight $s=1$. They can be expanded as
\begin{equation}
	f_s(\omega,\theta)=\sum_{\ell=|s|}^{\ell_{\max}}f_{\ell}{}^0(\omega)\,{}_sY_{\ell}{}^0(\theta)\,,
\end{equation}
where $f_s$ denotes any of the fields $\{{\bf K},{\bf k},{\widehat N}\}$. 

Since the free data $\aa, \bb, \NN, \kkappa, \Kc$ is exactly Schwarzschild, the parabolic-hyperbolic system takes the form
\begin{align}
	& \omega\partial_\omega\widehat{N} +\tfrac12\widehat{N}^2\,\eth\bar{\eth}\widehat{N}+\tfrac12\widehat{N} \nonumber \\ & \hskip0.7cm -\tfrac12\big[\omega^{-2}(\kkappa{\bf K} + \tfrac14{\bf K}^2) + (1-{\bf k}{\overline{\bf k}}) \,\big]\widehat{N}^3=0 \label{eq:schNh} \\
	& \omega\partial_\omega {\bf K}+2\,\big(\kkappa- \tfrac12{\bf K}\big)-\tfrac12\omega\widehat{N}\big[\,\bar{\eth}{\bf k}+\eth{\bar\kk}\, \big]\nonumber \\ & \hskip3.5cm -\omega\big[{\bf k}\,\bar{\eth}\widehat{N}+\overline{\bf k}\,\eth\widehat{N} \big]=0 \label{eq:schK} \\
	& \omega\partial_\omega {\bf k}-2\,{\bf k} -\omega^{-1}\Big[\tfrac12\widehat{N}\,{\eth}{\bf K} + \big(\kkappa - \tfrac12{\bf K}\big)\,{\eth}\widehat{N}\Big]=0\,. \label{eq:schk}
\end{align}
As $\widehat{N}$ vanishes at $\omega=0$, all the formally singular terms remain finite since $\widehat{N}\sim \omega$.

By virtue of
\begin{align}
	\eth\,{}_sY_{\ell}{}^m & = \sqrt{(\ell-s)(\ell+s+1)}\,\,{}_{s+1}Y_{\ell}{}^m\,,\\
	\bar{\eth}\,{}_sY_{\ell}{}^m & = -\sqrt{(\ell+s)(\ell-s+1)}\,\,{}_{s-1}Y_{\ell}{}^m\,,\\
	\bar{\eth}\eth\,{}_sY_{\ell}{}^m & =-(\ell-s)(\ell+s+1)\,\,{}_sY_{\ell}{}^m\,,
\end{align}
the Newman--Penrose $\eth$ and $\bar\eth$ derivatives act as algebraic operators on the spin-weighted spherical harmonics, which allows us to evaluate every angular derivative by an exact algebraic operation. In particular, the scalar Laplacian, relevant for the evaluation of the parabolic equation for ${\widehat N}$, acts as $\bar\eth\eth\,Y_{\ell}{}^m =-\ell(\ell+1)\,Y_{\ell}{}^m$. 

We treat the non-linearities by making use of the mode-coupling formulae of fully spectral methods (see, e.g., \cite{csukas2023possible}),
\begin{widetext}
	\begin{equation}
		({}_{s_2}f\cdot {}_{s_3}g)_{\ell_1}{}^{m_1} = \sum_{\ell_2=|s_2|}^{\ell_{\max}}\sum_{m_2=-\ell_2}^{\ell_2}\sum_{\ell_3=\max\{|s_3|,|\ell_1-\ell_2|\}}^{\ell_1+\ell_2}(-1)^{m_1+s_1}\,{}_{s_2}f_{\ell_2}^{m_2}\cdot {}_{s_3}g_{\ell_3}^{m_1-m_2}\times{}_{s_2}\,{}_{s_3}\,{}_{-s_1}G^{\,m_2\,\,m_1-m_2\,\,-m_1}_{\hskip0.05cm\ell_2\hskip0.6cm\ell_3\hskip0.67cm\ell_1}\,,
	\end{equation}
	where $s_1=s_2+s_3$ and ${}_{s_1}{}_{s_2}{}_{s_3}G^{m_1 m_2 m_3}_{\ell_1\ell_2\ell_3}$ stands for the modified Gaunt coefficient built from Wigner-$3j$ symbols as
	\begin{equation}
		{}_{s_1}{}_{s_2}{}_{s_3}G^{m_1 m_2 m_3}_{\ell_1\ell_2\ell_3}=\sqrt{\frac{(2\ell_1+1)(2\ell_2+1)(2\ell_3+1)}{4\pi}}
		\left(\begin{array}{ccc}
			\ell_1 & \ell_2 & \ell_3\\
			-s_1 & -s_2 & -s_3
		\end{array}\right)
		\left(\begin{array}{ccc}
			\ell_1 & \ell_2 & \ell_3\\
			m_1 & m_2 & m_3
		\end{array}\right)\,.
	\end{equation}	
\end{widetext} 
Note that this representation yields, in the Fuchsian equations \eqref{eq:Fuchsian1}, maximal generated $\ell$-modes $2\,\ell_{\max}$, $2\,\ell_{\max}$, and $5\,\ell_{\max}$ for ${\bf K}$, ${\bf k}$, and ${\widehat N}$, respectively. Nevertheless, we explicitly solve only for the modes with $\ell\le\ell_{\max}$, and merely monitor the coupling to the higher modes through the $L^2(S^2)$ norms.

\section{The norm used to monitor residuals}
The combined $L^2$-norm we use to monitor the residuals of the solutions to the Fuchsian equations \eqref{eq:Fuchsian1} is defined as
\begin{equation}\label{eq:complete-norm}
	\mathcal E_{\rm full}(\omega) =
	\left[\sum_{q\in\{{\bf K},{\bf k},{\widehat N}\}}\big\|\mathcal R_q(\omega)\big\|_{L^2(S^2)}^2\right]^{1/2}\,,
\end{equation}
where the partial $L^2$-norms are given by sums over the spherical modes,
\begin{equation}
	\big\|\mathcal R_q\big\|_{L^2(S^2)}^2 =\sum_{\ell\in S_q}\big|(\mathcal R_q)_{\ell}{}^0\big|^2\,.
\end{equation}
Note that, for the strictly polynomial equations, a degree-$\ell_{\max}$ expansion generates $\ell$-modes beyond $\ell_{\max}$ in the residuals. The norm $\mathcal E_{\rm full}$ is faithful in the sense that the sets $S_q$ comprise all of these modes, i.e., $\ell\leq 5\,\ell_{\max}$ for $\Nhat$ and $\ell\leq 2\,\ell_{\max}$ for $\KK,\kk$.

\section{Linearizing the parabolic equation}\label{sec:lin-par}
Denote by $\Nhat_{\rm mod}$ the lapse of the spherical companion with $E = 2$, $J =0$, determined by $\alpha$ and $\beta$ coefficients used in $\KK_1=\bg\KK_1+\alpha\,Y_0{}^0$ and $\Nhat_4=\bg\Nhat_4+\beta\,Y_0{}^0$. We use  $\Nhat_{\rm mod}$ as the background for estimating the amplification of angular momentum perturbations of $\Nhat$, we linearize \eqref{eq:schNh} around $\Nhat_{\rm mod}$. Substituting the expansion $\Nhat=\Nhat_{\rm mod} +\sum_{\ell=0}^{\ell_{\max}}\delta\Nhat_\ell{}^0\,Y_{\ell}{}^0$ and noticing that the $\ell$-modes decouple, we obtain from the principal part of \eqref{eq:schNh}
\begin{equation}
	\partial_\omega(\delta\Nhat_\ell{}^0)=\tfrac12\,\omega^{-1}\Nhat_{\rm mod}^2\,\ell(\ell+1) \,\delta\Nhat_\ell{}^0\,.
\end{equation}	
This implies for the individual modes
\begin{equation}
	\delta\Nhat_\ell{}^0=(\delta\Nhat_\ell{}^0)\big\vert_{\omega=0}\exp\bigg[ \tfrac12\,\ell(\ell+1) \int_{0}^{\omega}\widetilde{\omega}^{-1}\Nhat_{\rm mod}^2\,d\widetilde{\omega}\bigg]\,.
\end{equation}	
Therefore the exponent
\begin{equation}\label{eq:intGfk}
	\mathfrak{G}(\omega)=\tfrac12\,\ell(\ell+1) \int_{0}^{\omega}\widetilde{\omega}^{-1}\Nhat_{\rm mod}^2\,d\widetilde{\omega}
\end{equation}
is of critical importance in handling the propagation of numerical errors. Since numerical error always enters via a small but non-vanishing $(\delta\Nhat_\ell{}^0)\big\vert_{\omega=0}$, the above argument suggests that its growth can be kept below a threshold by arranging for the exponent $\mathfrak{G}$ to remain small. We chose the second hyperboloidal slice, relevant for these excitations, such that the integral $\int_{0}^{\omega}\widetilde\omega^{-1}\Nhat_{\rm mod}^2\,d\widetilde\omega$ in $\mathfrak{G}$ was approximately $0.18$ and, for the same reason, we used a conservative $80$-digit numerical precision. We believe that these choices, made in setting up our numerical method, played a crucial role in achieving the desired level of accuracy.

\section{Bondi energy and angular momentum}

For completeness, and also to correct two minor typos in (4.31) of \cite{csukas-racz-2025-hyperboloidal}, we record here the general form of the Bondi energy as a function of the free data and the asymptotic degrees of freedom $(\KK_1, \Nhat_4)$:
\begin{widetext}\vskip-0.2cm
	\begin{align}\label{eq:B-energy-ADF}
		& E_{B}= \frac{1}{32\pi}\int_{\partial\Sigma}\bigg[{\kkappa_0}^{-1}\KK_1^3+8 \,\kkappa_0^3 \,\Nhat_4-4\, \kkappa_0 \kkappa_3+12\, \aa_1 \kkappa_0{}^2-2\,\KK_1\big(3\,\kkappa_0{}^{-1}+3 \,\aa_0 \kkappa_0-4\, \kkappa_2\big)\bigg.\nonumber \\ 
		& -\Big\{\kkappa_0^5\,\NN_1^{-1/2}\,\ethb\big(\NN_1^{3/2}\kkappa_0^{-4}\KK_1\big)+2\,\kkappa_0\,\NN_2^2\,\ethb \left(\NN_2^{-1}\kkappa_0\right)+\tfrac{1}{4}\kkappa_0^5\,\eth\ethb\left(\kkappa_0^{-6}\KK_1\right)
		-\tfrac32\kkappa_0^{-2}\,\KK_1\,\eth\ethb\kkappa_0+\tfrac14{\kkappa_0}^{-1}{\eth\ethb\KK_1}+{\rm c.c.}\Big\}\bigg]\boldsymbol{\epsilon}_q\,.
	\end{align}
\end{widetext}\vskip-.8cm 
We also recall the Bondi angular momentum in terms of the asymptotic degree of freedom $\kk_2$
\begin{align}
	J_B[\phi] = -\frac{1}{16\pi}&\int_{\partial{\Sigma}}\Big\{(\phi^a\overline{q}_a)\,\kk_2 + (\phi^aq_a)\,\overline{\kk}_2\Big\}\,\boldsymbol{\epsilon}_q\,, \label{eq:Bondi-angmom-ADF}
\end{align}
where $\boldsymbol{\epsilon}_q=\sin\vartheta\,\mathrm{d}\vartheta\wedge\mathrm{d}\varphi$ is the volume form of the center-of-mass unit sphere metric, and $\phi^a$ is the axial vector field maximizing the functional \cite{Racz:2024giv}.


\end{document}